\documentclass[twocolumn,noshowpacs,nofootinbib,colorlinks,hyperindex,unicode,10pt]{revtex4-2}
\usepackage{graphicx}
\usepackage{tikz}

\usepackage[compat=1.1.0]{tikz-feynman}

\newcommand\orcidg{{\href{https://orcid.org/0000-0002-7942-7941}{\orcidicon}}}
\newcommand\orcidNog{{\href{https://orcid.org/0000-0002-1827-1031}{\orcidicon}}}

\usepackage{amsfonts,amsmath,amssymb,tikz,accents}
\usepackage[eulergreek]{sansmath}
\usepackage{indentfirst,bigints}
\usepackage{hyperref,upgreek}
\usepackage{bm,graphics,color,xcolor}
\usepackage{feynmp-auto}
\usepackage{slashed}
\counterwithout{equation}{section}
\usepackage{hyperref}
\usepackage{pgfplots}
\pgfplotsset{compat=1.16}

\hypersetup{hidelinks,backref=true,pagebackref=true,hyperindex=true,colorlinks=true,breaklinks=true,urlcolor= blue}
\hypersetup{%
  colorlinks = true,
  linkcolor  = blue,
  citecolor = cyan,
}
\usepackage{textgreek}
\usepackage{color,xcolor}

\def\0{\mbox{\boldmath$\displaystyle\mathbb{O}$}}

\def\p{\partial}

\newcommand\orcidroldao{{\href{https://orcid.org/0000-0003-3978-532X}{\orcidicon}}}
\newcommand{\orcidicon}{%
	\begin{tikzpicture}
	\draw[lime, fill=lime] (0,0)
		circle [radius=0.16]
		node[white] {{\fontfamily{qag}\selectfont \tiny ID}};
	\draw[white, fill=white] (-0.0625,0.095)
		circle [radius=0.007];
	\end{tikzpicture}	\hspace{-2mm}
}

\newcommand\orcidMa{{\href{ https://orcid.org/0000-0002-7888-3996}{\orcidicon}}}

\usepackage{float,xcolor,upgreek}
\usepackage{color} 
\usepackage{tikz-feynman}
\tikzfeynmanset{compat=1.0.0}
\newcommand{\beq}{\begin{eqnarray}}
\newcommand{\eeq}{\end{eqnarray}}
\newcommand{\bea}{\begin{eqnarray}}
\newcommand{\eea}{\end{eqnarray}}
\DeclareMathOperator{\sign}{sign}

\begin{document}

\title{ Topological insulator realization induced by fermionic interaction through BF mediators     }

\author{G. B. de Gracia \orcidg{}}
\affiliation{Federal University of Triângulo Mineiro, Physics department, 
38064-200, Uberaba, MG, Brazil}
\email{gabriel.gracia@uftm.edu.br}
\author{R. da Rocha \orcidroldao\!\!}
\affiliation{Federal University of ABC, Center of Mathematics,  Santo Andr\'e 09210-580, Brazil.}
\email{roldao.rocha@ufabc.edu.br}
\author{A. A. Nogueira \orcidNog{}}
\affiliation{Santa Catarina State University, Department of Physics, Joinville, 89219-710, SC, Brazil.}
\email{andsogueira@hotmail.com}
\author{M. A. C. de Barcelos\orcidMa}
\affiliation{Federal University of Triângulo Mineiro, Physics department, 
38064-200, Uberaba, MG, Brazil }
\email{d202010309@uftm.edu.br}


\begin{abstract} 
\indent This paper demonstrates that ordinary fermions, interacting via a BF mediator, form a renormalized structure with non-trivial topological properties.  We explore the analogy between the renormalized fermion and a subset of real
three-dimensional topological insulators (TIs) in the vicinity of their single Dirac cones.  Using the typical magnitude of TI lattice and gaps, we constrain model parameters.  The topological structure induced by radiative corrections enters a class of modified Dirac equations, ensuring the existence of helical gapless near-boundary modes. We derive an effective potential for inter-quasi-particle interactions that includes finite-size effects in the axial direction and reveals spin-orbit coupling and time-reversal invariance signatures. Remarkably, particles with different spins do not interact via the characteristic spin-orbit coupling, and a non-central spin-dependent force arises. We also address the behavior of the system in the phase transition associated with the thin film limit.

\end{abstract}

\maketitle


\section{Introduction}

\indent The issue of topological excitations in condensed matter systems reveals a rich research field that benefits from an intense interplay between theory and experiment. Interestingly, in the 1980s, several developments predicted the possible existence of three-dimensional topological insulators (TIs)  \cite{quatro}. Later, theoretical studies evinced the possibility of TIs mainly during the first decade of the $21$st century \cite{um,dois,tres}, highlighting the fundamental role of time-reversal and parity invariance. Also, a set of experimental advances indeed revealed the existence of a three-dimensional TI \cite{cinco,seis}. Regarding TIs such as Bi$_2$Se$_3$ and Bi$_2$Te$_3$, their theoretical characterization was developed in the seminal papers \cite{sete, oito, nove}. The system exhibits a single Dirac point near the $\Gamma=(0,0,0)$ high-symmetry point of the reciprocal lattice, displaying a quasi-particle structure in this regime.\\
\indent Since the non-trivial topological nature implies robust boundary modes, one can suggest their use as qubits in quantum computing, a topic of recent interest \cite{dez, qb1, qb2}. Moreover, these modes possess a helical character, whereby the spin orientation is intrinsically locked to the direction of propagation. Then, the manipulation of this degree of freedom gave rise to spintronics \cite{onze,doze}. In this context, the theoretical and experimental implementation of spin filters to select these modes plays an important role in technology.  Recently, relevant research on the theoretical description of Bi$_2$Se$_3$ has been devoted to modeling new varieties of diodes \cite{treze} and the realization of the Hall effect \cite{quatorze}, for example. There are also several other recent examples of the interplay between theoretical physics methods and condensed matter research, including quantum field theory inputs,  and correlated boundary effects \cite{1,3,5,8,9,10,daRocha:2023waq,DaRocha:2020oju,Ferreira-Martins:2019wym}. 

\indent Therefore, we aim to further develop the fundamentals of TIs, demonstrating that ordinary fermions interacting via a specific formulation of BF-like bosonic mediators lead to a renormalized system with non-trivial topological properties. A variety of analogies with the overall phenomenology of 3D TIs near the Dirac cone are implemented. The radiative corrections are carefully analyzed, allowing the determination of parameter ranges compatible with the topological nature. In this setup, our model represents the long-range, continuous description of a subjacent lattice system, defined up to a cutoff scale. Throughout the paper, inputs from real 3D TIs include typical orders of magnitude to fix our field-theory parameters. We {also highlight that the emerging renormalized fermion has a topological structure that ensures gapless helical boundary states. It is characteristic of a whole class of modified Dirac fermions \cite{shen,B1}. These results are used to derive an inter-quasi-particle potential. Remarkably, its structure is such that the ones with different spins do not interact through the characteristic spin-orbit term. The latter implies spin-dependent torques whose equilibrium is achieved when both particles reach the same momentum orientation.} 
Since TIs have finite sizes in the real world, we consider compactification in the axial direction and evaluate the transition from the large-width setup to the thin-film limit. In the second case, there are configurations in which the characteristic spin-dependent force is eliminated. Interestingly, in experimental realizations, the transition between these two mathematical regimes is achieved by varying the sample width by just a few nanometers \cite{quinze}.

Given the potential technological applications, we emphasize that the possibility of inducing and controlling quasi-particle properties analogous to those of real three-dimensional TIs is of particular interest. Such control may enable technological advances that extend beyond the original scope of our field-theoretical investigation of topological systems. 

\indent The paper is organized as follows. In Sec. \ref{2}, the model is composed of fermions interacting via a current and spin pseudo-tensor through a four-dimensional BF mediator. The role of Lagrange multipliers, finite-size effects, and the obtainment of the propagator are presented. Sec. \ref{3} is devoted to demonstrating that the renormalized fermionic structure evinces a non-trivial topological nature owing to the specific interaction mediated by the BF bosonic sector. Then, in Sec. \ref{4}, the potential interaction between the boundary modes protected by the renormalized structure is derived, revealing the characteristic spin-orbit interaction and an overall structure compatible with the physics of helical states. The transition to the thin-film limit is also investigated, and comparisons with standard phenomenology are derived. In Sec. \ref{5}, we conclude. Natural units are used throughout $c=\hbar=1$.

\section{The model, its Lagrange multipliers, boundary structure, and propagators \label{2}}

\begin{figure}[H]
	\centering
\includegraphics[width=8cm]{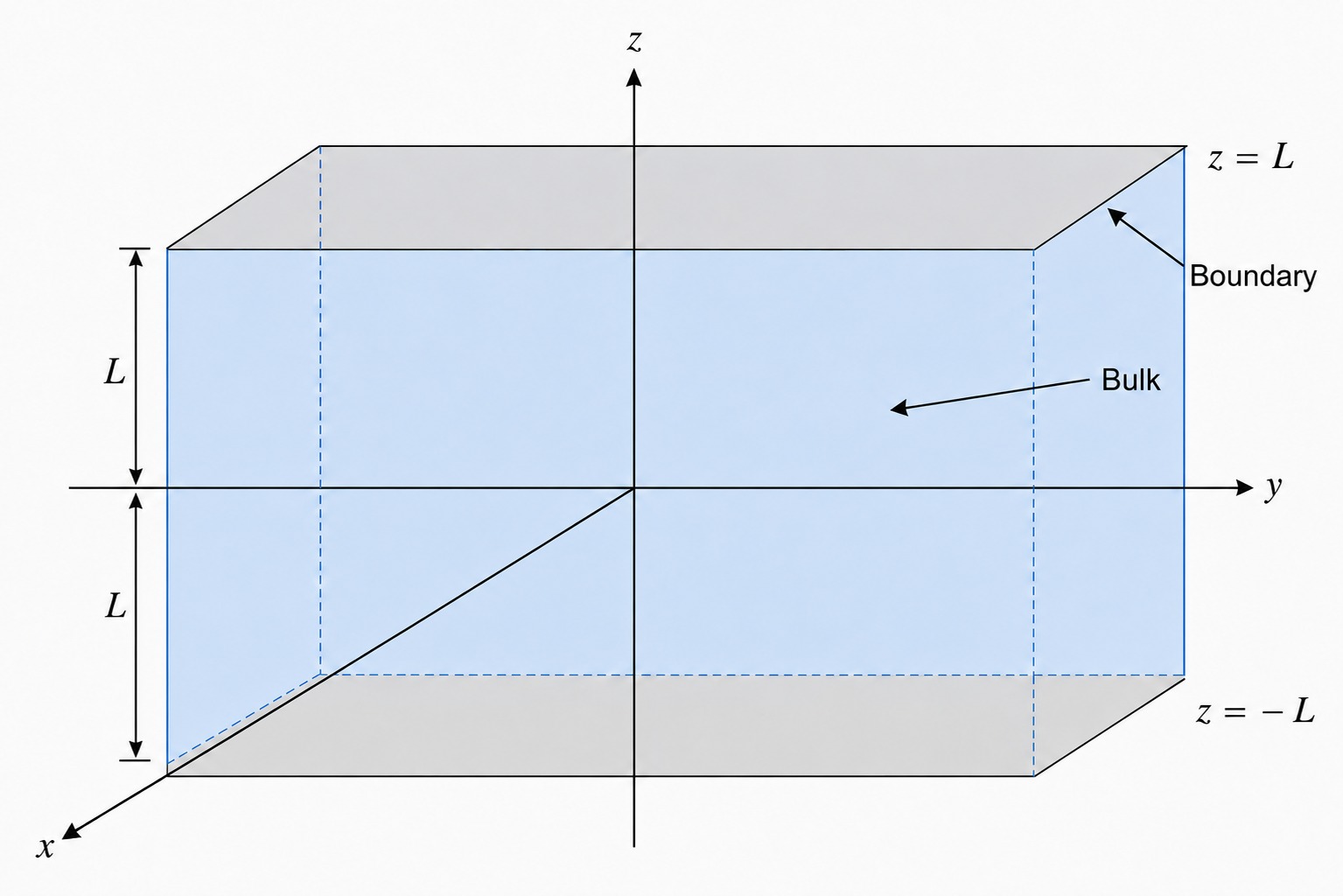}
	\caption{ The figure highlights our general setup. In our approximation, we consider a compactification in the axial coordinate, with the remaining dimensions infinite. The field theory is defined within the bulk and its boundaries. }
	\label{fig1}
\end{figure}

\indent In this section, we introduce fermionic fields interacting by means of a BF model of intermediate bosons. We will show that the renormalized fermion inherits a topological nature induced by this specific interaction. The equations of motion resemble the superposition of two Hall-like effects for both spin and currents. The vector field represents the effective electromagnetic interaction, whereas the pseudo-tensor one accounts for a spin interaction. Unlike quantum Hall phenomenology, this system has parity and time-reversal symmetry, which is compatible with a TI description. This can be ensured if the rank-two field transforms as a pseudo-tensor.  \\
\indent Beyond the aforementioned helical states arising in TIs, there is a wide variety of interesting boundary phenomena in condensed matter, with potential technological applications. It is interesting to mention some recent field-theoretical approaches to modeling them \cite{8,9,10}. To include finite-size effects, we consider the TI confined between two infinite planes whose normal lies in the $x_3$ direction. {According to Figure \ref{fig1}, the bulk region, as well as its defining boundaries, are explicitly illustrated}.
The action is given by 
\begin{eqnarray}
       S\!&\!=\!&\!\int d^4x\Big[ \varepsilon_{\mu \nu \alpha \beta} \mathcal{B}^{\mu \nu}(x) \p^\alpha A^\beta(x)+C(x)\p_\mu A^\mu(x)\nonumber\\&&
       \quad\quad+\chi^\gamma(x) \p^\nu \mathcal{B}_{\nu \gamma}(x)+\sigma(x) \p_\gamma \chi^\gamma(x)+A_\mu(x) J^\mu(x) \nonumber \\ 
       && \quad\quad +\mathcal{B}_{\mu \nu}(x)T_{\mu \nu}(x)+i\bar{\Psi}\slashed{\partial}\Psi(x)-m\bar{\Psi}(x)\Psi(x) \Big]                \end{eqnarray}
      with the fermionic sources associated with the electromagnetic current density, with $e$ denoting the electric charge, $J^\mu(x)=e\bar{\Psi}(x)\gamma^\mu \Psi(x)$, and a spin sector represented by the pseudo-tensor $T_{\mu \nu}(x)= \frac{g}{4}\bar{ \Psi}(x)\gamma_5\left[\gamma^\nu,\gamma^\mu\right]\Psi(x)$.
      The first-order structure of the model is compatible with using periodic boundary conditions for the fields to ensure differentiability. Then\footnote{Here $x_{\scalebox{0.6}{$\parallel$}}$ denotes space-time coordinates different from $x_3$.}, $\Phi(x_3=L, x_{\scalebox{0.6}{$\parallel$}})=\Phi(x_3=-L,x_{\scalebox{0.6}{$\parallel$}})$, for all fields denoted by the collective  notation $\Phi(x)$, considering the boundaries located at $x_{3 \pm}=\pm L$. Since the action is adimensional in natural units, the pseudo-tensor field has mass dimension $2$. The vector and fermion fields display their known dimensions, whereas the ones from the auxiliary fields are chosen appropriately.   Clearly, $g$ is a dimensionful coupling constant. Although it indicates a non-renormalizable theory, we are interested in an effective continuous description of a microscopic lattice system at large scales. Then, the cutoff is physical and defined by the inverse of a typical inter-site distance. This approach suffices to derive relevant conclusions regarding the connection between topology and the one-loop fermionic mass correction in the next section.
      
       The invariance under gauge symmetry transformations from the free action without sources\footnote{The space-time derivatives of the gauge parameters $\Lambda_\nu(x)$ and $\lambda(x)$ are assumed to be periodic.}
       \bea \!\!\!\!\!\delta \mathcal{B}_{\mu \nu}(x)=\partial_\mu \Lambda_\nu(x)-\partial_\nu \Lambda_\mu(x),  \quad \delta A_\mu(x)=\partial_\mu \lambda(x), \eea
       is partially broken due to the form of the tensor source that couples with $\mathcal{B}_{\mu \nu}(x)$. However, to invert the theory's differential operator and derive the propagator, one must impose conditions implied by the auxiliary fields set $C(x)$, $\chi^\gamma(x)$, and $\sigma(x)$. An alternative way to analyze the situation is to consider the canonical approach \cite{naka}. The auxiliary fields must be introduced to furnish a second-class system from the beginning, implied by the primary momentum definition. This procedure establishes a well-defined structure for the associated Dirac brackets, allowing quantization by the correspondence principle in an indefinite metric Hilbert space. It is worth mentioning that, in this case, the specific structure of the gauge condition may affect the physics since conditions resembling $\alpha$-like gauges can lead to longitudinal contributions to the boson propagator that are not canceled out by attaching the $T_{\mu \nu}(x)$ pseudo-tensor in the expression of a given amplitude. However, owing to the gauge freedom of the free action, one readily concludes that this choice is appropriate to eliminate spurious longitudinal non-physical sectors by coupling the sources only with the physical transverse parts of the intermediate bosonic fields. It is important to state that this is not a gauge fixing procedure. Instead, it comprises establishing an entirely new interacting theory defined by the Lagrange multiplier constraints.
       It is also possible to conclude that they lead to functional delta functions, implying conditions on all quantum paths, excluding the contributions of the longitudinal source sectors in all physical observables.  
       
        Additionally, one can also consider the equivalent path integral formulation.  The tensor source has the following general tensor structure written in terms of vector fields and pseudo-vector fields
       \bea\label{eq10}  T_{\mu \nu}=\epsilon_{\mu \nu \gamma \beta}\p^\gamma \mathcal{H}^\beta +\p_\mu \mathcal{A}_\nu-\p_\nu \mathcal{A}_\mu. \eea
Noting that just the transverse part of the pseudovector $\mathcal{A}_\mu $ contributes to Eq. (\ref{eq10}), the longitudinal part of $T_{\mu \nu}$ can be eliminated by a suitable change of variables with unit Jacobian associated to the auxiliary field $\chi_\gamma(x) \mapsto \chi_\gamma(x)+g\mathcal{A}_\gamma^\intercal(x)$ with ${\scalebox{0.8}{$\intercal$}}$  denoting the transverse component  \cite{kr}. To summarize, the complete structures of all connected Green functions are not affected by the longitudinal sector of the sources. The associated propagator structure ensured by our choice of gauge condition is projected out in the transverse sector, in all orders of approximation. The latter is determined only by the transverse sector of the tensor interactions. Therefore, although symmetry-breaking terms may be induced by radiative corrections in the quantum action, the presence of the auxiliary fields avoids their contribution to physical processes.

Considering this specific interaction, it is possible to show that using the Feynman rules introduced in the next section leads to the following one-loop polarization tensor for the mixed bosonic two-point function:
\bea \Pi_{\mu \alpha \beta}(k)=2i (eg)\epsilon_{\mu \nu \alpha \beta}k^\nu \left(\frac{2}{\varepsilon}\right),\eea
where only the divergent part\footnote{Considering the limit $\epsilon \to 0$ in dimensional regularization.} is  highlighted in the dimensional regularization scheme. Moreover, the $L\to \infty$ limit was taken into account, for simplicity. Regarding the bosonic two-point part, there is a QED-like rank-2 polarization tensor for the vector field and also a rank-4 one for the tensor field. Therefore, considering the quantum action and the fact that the cutoff is physical and finite, defined by the continuum limit of the underlying lattice system, no issues arise. In fact, throughout this paper, and complying with its original conception,  the renormalization conditions are not a method to absorb infinities but a prescription to constrain the model to meet a series of physical demands, the so-called on-shell renormalization conditions.

\subsection{Obtaining the propagator}

\indent The integration region can be split as       
       \bea \int d^4x\equiv \int_{-L}^L dx_3 \int d^3x  \eea
The imposed periodicity replaces momentum integrals with sums
    \bea \int \frac{dp_3}{(2\pi)}\mapsto \sum_n \frac{1}{2L} \eea
since, in this case, the allowed set of $p_3$ momenta is discrete $p_3=\frac{\pi n}{L}$, with $n$ being an integer. \\
\indent The zeroth-order propagator, including the sum associated with the finite-size effects in the $x_3$ direction, as well as the auxiliary field constraints, reads
\begin{widetext}\indent 
\bea G_{\nu \alpha \beta}(x_{\scalebox{0.6}{$\parallel$}},x_3)=\frac{i}{2L}\sum_n\int \frac{d^3p_{\scalebox{0.6}{$\parallel$}}}{(2\pi)^3}\frac{e^{i\left(-p^\mu_{\scalebox{0.6}{$\parallel$}} x_{\mu \parallel}+\frac{\pi n}{L} x_3 \right)}\varepsilon_{\mu \nu \alpha \beta}\tilde{p}^\mu  }{\big(p^2_{\scalebox{0.6}{$\parallel$}}-\frac{\pi^2n^2}{L^2}\big)},            \label{prop} \eea
\end{widetext}
with $\tilde{p}^\mu=(p_{\scalebox{0.6}{$\parallel$}}^\mu, \frac{\pi n}{L})$ and $p^2_{\scalebox{0.6}{$\parallel$}}=p_0^2-|\vec p_{\scalebox{0.6}{$\parallel$}}|^2$. The symbol $\parallel$ denotes a quantity evaluated on surfaces whose normal lies in the $x_3$-direction. By inspecting the propagator structure, one immediately confirms that the suitable introduction of the Lagrange multipliers indeed ensures transversality. In the next section, we will prove that it also implies that just the transverse piece of the vertices contributes to the fermionic self-energy.

In the static limit, a useful expression to derive the effective interaction potential is obtained
\begin{widetext}
\bea G_{\nu \alpha \beta}(\vec{x}_{\scalebox{0.6}{$\parallel$}},x_3)=\int \frac{d^2p_{\scalebox{0.6}{$\parallel$}}}{(2\pi)^2}\frac{e^{\left(i\vec{p}_{\scalebox{0.6}{$\parallel$}}\cdot\vec{x}_{ \parallel}-\sqrt{\vec{p}^2_{\scalebox{0.6}{$\parallel$}}}|x_3| \right)}\coth{(L\sqrt{\vec{p}^{\,2}_{\scalebox{0.6}{$\parallel$}}})}\varepsilon_{\mu \nu \alpha \beta}\tilde{\vec{p}}^{\,\mu}  }{2i\sqrt{\vec{p}^{\,2}_{\scalebox{0.6}{$\parallel$}}}},    \label{prop2}         \eea
\end{widetext}
with the definition $\tilde {{\vec{p}}}^{\,\mu}=(0,\vec{p}_{\scalebox{0.6}{$\parallel$}},i\sign x_3 |\vec{p}_{\scalebox{0.6}{$\parallel$}}|)$. 
In the $L \to \infty$ limit, the axial momentum approaches the continuum, the sum tends to an integral and, instead of equation \eqref{prop}, the full propagator  becomes 
\bea  G_{\nu \alpha \beta}(x_{\scalebox{0.6}{$\parallel$}},x_3)=i\int \frac{d^4p}{(2\pi)^4}\frac{\varepsilon_{\mu \nu \alpha \beta}\ p^\mu e^{-ip^\mu x_\mu}}{p^2}.     \label{largeprop}          \eea

Now, let us define a set of approximations for the static propagator \eqref{prop2}.
Since $\coth(x)\approx \frac{1}{x}+\frac{x}{3}$ for $0\leq x\lesssim 1$, this approximation is valid for $0\leq|\vec{p}_{\scalebox{0.6}{$\parallel$}}|\lesssim \frac{1}{L}$. 
Therefore, in the thin-film limit $L\ll 1$, one can replace the integration limit  $0\leq|\vec{p}_{\scalebox{0.6}{$\parallel$}}|< \infty$ and utilize the Taylor expansion in the static propagator expression. The second term of the expansion is not considered in practice, since it would lead to delta-like contributions when computing interparticle interaction potentials. We put this kind of contribution aside since our approach is based on a continuous long-distance description of an underlying lattice system. As we shall see, the topological nature induced by the BF interaction implies that charge carriers are restricted to move on the boundaries. Then, we impose this constraint at sources and analyze the associated phenomenology. Therefore, for particles interacting on the same boundary, one has
\bea G_{\nu \alpha \beta}(\vec{x}_{\scalebox{0.6}{$\parallel$}},0)\!=\!\int\! \frac{d^2p_{\scalebox{0.6}{$\parallel$}}}{(2\pi)^2}\frac{e^{(-i\vec{p}_{\scalebox{0.6}{$\parallel$}}\cdot\vec{x}_{ \parallel})}\varepsilon_{\mu \nu \alpha \beta}\tilde{\vec{p}}^{\,\mu}  }{2iL\vec{p}^{\,2}_{\scalebox{0.6}{$\parallel$}}}.           \label{prop3}  \eea

On the other hand, quasi-particles at different surfaces $|x_3|=2L$, in the thin film approximation, interact by means of the following propagator
\bea \!\!\!\!\!\!\!\!\!G_{\nu \alpha \beta}(\vec{x}_{\scalebox{0.6}{$\parallel$}},\pm 2L)\!=\!\!\int \!\!\!\frac{d^2p_{\scalebox{0.6}{$\parallel$}}}{(2\pi)^2}\frac{e^{(-i\vec{p}_{\scalebox{0.6}{$\parallel$}}\cdot\vec{x}_{ \parallel})}\varepsilon_{\mu \nu \alpha \beta}\tilde{\vec{p}}^{\,\mu}(1\!-\!2L|\vec{p_{\scalebox{0.6}{$\parallel$}}}|  )  }{2iL\vec{p}^{\,2}_{\scalebox{0.6}{$\parallel$}}}   \label{prop4}           \eea
considering the approximation for the exponential term.

\section{On the radiative corrections \label{3}}

\indent Our objective throughout this section is to explicitly demonstrate that this specific coupling with the BF mediator fields can indeed induce a non-trivial topological system by the renormalized fermionic structure. It is worth mentioning that there are correlated studies in condensed matter using this mediator \cite{bf}, although focused on bosonization instead of the renormalized fermionic response as in our present research. In this context, one can also cite the recent paper on topological fermion properties induced by the Chern-Simons interaction in honeycomb lattices \cite{gab}. \\
 \indent In this specific research, the cutoff is physically motivated, being associated with the inverse of a typical three-dimensional  TI length scale, beyond which the continuous description would not be appropriate. At this stage, finite-size effects are disregarded due to the preliminary nature of this investigation on the overall topological properties.  
 
To derive the fermion self-energy, let us establish the Feynman rules. The structure of the vertices for the electromagnetic sector is given as follows:
\bea -ie\gamma_\mu,                \eea
whereas for the spin sector, it reads
\bea  -\frac{ig}{4}[\gamma_\mu,\gamma_\nu]\gamma_5.                                  \eea
\noindent Considering the auxiliary Lagrangian sector as well as the specific structure of the couplings, one obtains the mixed gauge field Feynman propagator\footnote{With the additional $i$ factor to be accounted in the loops.}  
\bea  G^{\beta \mu \nu}(p)=-\frac{\varepsilon^{\chi \beta \mu \nu}p_\chi}{p^2},                                            \eea
which couples only with the transverse part of the interaction vertices. Then, according to previous discussions, we explicitly verify that longitudinal source components do not enter either loops or in the interaction potential computations. Throughout this section, we consider the system in the large-width regime, associated with Eq.  \eqref{largeprop}. The thin-film limit is treated separately at the end, where we discuss the associated cancellation of topological character at the transition. \\
\indent The fermionic propagator reads
\bea   S=\frac{i(\slashed{p}+m)}{p^2-m^2}.                        \eea

 The induced fermionic one-loop self-energy correction can be managed to yield\footnote{Although the system is not renormalizable, it is just an effective model to describe a class of systems that indeed present physical cutoffs.}
\bea \!\!\!\!\!\!\!\!\!i\Sigma(p)\!=\!-\frac{ieg}{2}\!\!\!\int\!\! \frac{d^4k}{(2\pi)^4}\frac{\gamma_\beta  \varepsilon^{\chi \beta \mu \nu}(p\!-\!k)_\chi \gamma_\mu \gamma_\nu \gamma_5(m\!-\!\slashed{k})  }{(k-p)^2[k^2-m^2]}.                    \eea
Using the Feynman parametrization, it can be rewritten as
\bea \!\!\!\!\!\!\!\!i\Sigma(p)\!=\!-3eg \!\int\! dx\! \int\! \frac{d^4k}{(2\pi)^4}\frac{\big(p^2x(1\!-\!x)\!-\!m\slashed{p}x\!-\!k^2\big) }{[k^2-\Delta(p)]^2},                    \eea
with $\Delta=m^2x-p^2x(1-x)$. Then, we are led to
\begin{widetext}
\bea i\Sigma(p)=-\frac{3ieg}{16\pi^2} \int dx \left[ \frac{1}{\epsilon}+\ln{\frac{\tilde{\mu}^2}{\Delta}} \right]\left[
3p^2(1-x)x-m\slashed{p}x-2m^2x\right].                   \eea
\end{widetext}
\indent By replacing the theory parameters, considering the wave function and mass renormalization, one derives the following structure with the inclusion of counterterms
 \bea  \Sigma^R(p)=\Sigma(p)+D\slashed{p}+(C-Dm_R),                                \eea
 determined by the physical demand of a pole in the renormalized mass $m_R$ with unit residue
 \bea \Sigma(m_R)+C=0\ , \quad\quad \frac{d \Sigma(p)}{dp^2} \Big\vert_{p^2=m^2_R}+D=0,     \eea
 defining the so-called on shell renormalization conditions.\\
 \indent We express our dimensionally regulated integrals in terms of a physical cutoff 
 \beq \frac{1}{\epsilon}+\ln{\frac{\tilde{\mu}^2}{\Delta}} = \ln{\frac{\Lambda^2}{\Delta}}, \eea using the Veltman--Passarino procedure $\Lambda^2=\tilde \mu^2e^{1/\epsilon}$. We assume a finite cutoff $\approx \frac{12}{a^2}$, associated with a physical length $a$ related to the underlying lattice. Inspired by real three-dimensional  TIs such as Bi$_2$Se$_3$ \cite{sete,oito}, we consider $a\approx 30\, \text{\r{A}}$, which has the same order as the largest primitive lattice vectors connecting two equivalent sites in the material's characteristic quintuple layers. \\
 \indent The explicit expression for the fermion self-energy has three contributions
\begin{widetext}\begin{eqnarray}\label{eq20}\!\!\!\! \Sigma^{p^2}(p)\!&=&\!\frac{eg}{32\pi^2 p^4}\Bigg[ 
  p^2\left(6m^4_R\!-\!6m^2p^2\!-\!p^4\left(3\ln{\frac{\Lambda^2}{m^2_R}}\!+\!5\right)       \right) \!-\!3\left(2m^6_R\!-\!3m^4_Rp^2\!+\!p^6   \right)\ln{\frac{m^2_R}{(m^2_R\!-\!p^2)}}              \Bigg],\\
\Sigma^{m^2}(p)&=&\frac{3m^2_R eg}{8\pi^2 p^4}\Bigg[-p^2\left(-p^2\ln{\frac{\Lambda^2}{m^2_R}}+m^2_R-2p^2\right)   +(m^2_R-p^2)^2\ln{\frac{m^2_R}{(m^2_R- p^2)}}              \Bigg],\\
\label{eq40}\Sigma^{m_R \slashed{p}}(p)&=&\frac{3\slashed{p}m_R\ eg}{16\pi^2  p^4}\Bigg[-p^2\left( - p^2\ln{\frac{\Lambda^2}{m^2_R}}+m^2_R-2p^2\right)   +(m^2_R-p^2)^2\ln{\frac{m^2_R}{(m^2_R-p^2)}}              \Bigg].  \end{eqnarray}\end{widetext}
\indent Eqs. (\ref{eq20}) -- (\ref{eq40}) clearly reveal the quasi-particle creation threshold at $p^2>m^2_R$, leading to an imaginary contribution for $\Sigma(p)$ which implies a finite lifetime for the quasi-particle in such a configuration. However, near the renormalized mass shell, $p^2\approx m^2_R$, the imaginary part vanishes, leading to a long-lived renormalized system in the quadratic approximation. This kind of quadratic truncation of a more complex system is commonly used in topological condensed matter systems, considering a limited range of the Brillouin zone around a high-symmetry point or Dirac cone. In our case, we also expand the fermionic Lagrangian near the minimal gap defined by the renormalized mass, leading to
\begin{eqnarray}
    \Sigma(p^2)&\!=\!&-\frac{eg}{32\pi^2}p^2\left(3\ln{\frac{\Lambda^2}{m^2_R}}\!+\!5\right)\!\!+\frac{3m^2_R eg }{8\pi^2}\left(\ln{\frac{\Lambda^2}{m^2_R}}+1   \right)\nonumber\\&&+\frac{3\slashed{p}m_R\ eg}{16\pi^2}\left(\ln{\frac{\Lambda^2}{m^2_R}}+1  \right).               \end{eqnarray} 
In this approximation, the on-shell renormalization conditions fix the counterterms as $D=\frac{egm_R}{8\pi^2}$ and $C=-eg\Big[\frac{15m^2_R}{32\pi^2}\ln{\frac{\Lambda^2}{m^2_R}}+\frac{13m^2_R}{32\pi^2} \Big]$.
\indent The effective fermionic dispersion becomes 
\begin{widetext}
\bea 
\left[\slashed{p}
\left(1+\frac{3e g \ m_R}{16\pi^2}
\left(\ln\frac{\Lambda^2}{m^2_R} +\frac{5}{3}  \right) \right) -\left(m_{\textit{eff}}-eg\frac{\vec{p}^2}{32\pi^2}\left(3\ln{\frac{\Lambda^2}{m^2_R}}+5  \right)    \right)\right] \Psi=0,  \label{renorm}   \eea
\end{widetext}
with the definition of the energy-dependent effective mass\footnote{Although it is a useful concept, the pole is located at $m^2_R$, as already mentioned. }
\begin{widetext}
\bea
m_{\textit{eff}}(E)=m_R+eg\left( 3\frac{m^2_R}{32\pi^2}\ln{\frac{\Lambda^2}{m^2_R}}+\frac{5m^2_R}{32\pi^2}+\frac{\ E^2}{32\pi^2}\left(3\ln{\frac{\Lambda^2}{m^2_R}}+5\right)\right)  .\eea\end{widetext}
Although the system is not renormalizable, it can furnish an interesting effective model since one can establish it as a continuous physical limit of an underlying lattice system. Considering this feature, the previous discussion on the order of magnitude for the length scale $a$, and the mass gap as being $\approx 0.3$ eV, inspired by the order of magnitude from a set of known three-dimensional  TIs, one concludes that $\ln{\frac{\Lambda^2}{m^2_R}}\approx 13.3$.\\
\indent According to Refs.  \cite{wang1,wang2}, the topological properties must be evaluated by considering the so-called topological Hamiltonian in a zero-frequency surface $H_T\equiv H_0-\gamma_0\Sigma(\vec p,p_0=0)$. 
{At this stage, it is important to address the role of Lorentz symmetry. The Lorentz-invariant formulation adopted in the loop calculation is introduced as a technical simplification and is not intended to represent an exact symmetry of the condensed-matter system. In realistic topological insulators, the relevant quasiparticle velocity $V$ is much smaller than the speed of light. However, the topological invariant is determined by the global structure of the topological Hamiltonian and remains unchanged under smooth deformations of the effective theory, provided that the bulk gap does not close. Therefore, Lorentz- and rotational-symmetry-breaking corrections associated with the physical value of $V$ are expected to modify quantitative parameters of the dispersion relation while preserving the topological classification as long as the renormalized gap remains finite.}

Additionally, for systems with time reversal and parity invariance, possessing a topological Hamiltonian with a four-dimensional matrix structure \cite{dois}, the associated topological invariant is given by the product of the signs of the momentum-dependent gaps from all the filled bands evaluated at the time reversal invariant points 
 \bea  (-1)^\nu=\Pi_i\,\textsc{Sign}(M(p_i,0)),          \eea
 with $M(p)$ being a notation for the energy-momentum dependent gap. In our case, for computing topological numbers,  
 \begin{equation}
M(\vec p,0)=m_{\textit{eff}}(E=0)-eg\frac{\vec{p}^{\,2}}{32\pi^2}\left(3\ln{\frac{\Lambda^2}{m^2}}+5\right).\end{equation} \\If $\nu=1$, the system presents a non-trivial topology, defining the condition for the existence of robust boundary states. {However, for our continuous model, the topological nature is equivalently assigned if $\lim_{p^2\to \Lambda^2}\textsc{Sign}(M(p^2,0))=-\textsc{Sign}(M(0,0))$ }. It accounts for the spin orientation variation of the system evaluated at zero and at the cutoff momentum.
Therefore, the system is topological if 
 $g\geq 3\times 10^{-4}\; {eV}^{-1}$. {In order to evaluate the influence of the cutoff parameter, defining the macroscopic continuum picture of the underlying lattice, one could also consider the inverse of the Bi$_2$Te$_3$ largest primitive vector, which also belongs to the class of systems studied here. Interestingly, for Bi$_2$Te$_3$, it reads $30.5\text{\r{A}}$ while for Bi$_2$Se$_3$ the value is $29\text{\r{A}}$. Considering our choice for the gap size, one notices that for a cutoff $\Lambda$ higher than the gap, the expression for the momentum-dependent mass clearly shows that increasing $\Lambda$ yields a smaller lower bound for the coupling $g$, defining the topological range. Then, although considering a  Bi$_2$Te$_3$-like cutoff reduces this lower bound, {it implies a negligible modification if one considers the perturbative regime}. {To quantify this statement, let us compare the cutoff scales associated with
Bi$_2$Se$_3$ and Bi$_2$Te$_3$. Taking the corresponding largest primitive
lattice vectors, $a_{\rm Se}\approx29\,\text{\AA}$ and
$a_{\rm Te}\approx30.5\,\text{\AA}$, one finds
\bea
\frac{\Lambda_{\rm Te}^2}{\Lambda_{\rm Se}^2}
=\left(\frac{29}{30.5}\right)^2\simeq0.904,
\eea
which corresponds to an approximately $10\%$ reduction in the cutoff.
Since the lower bound of the coupling implied by the topological criterion depends only logarithmically on $\Lambda$, and considering the perturbative regime, it is approximately  proportional to the inverse of
\beq
3\ln\!\left(\frac{\Lambda^2}{m_R^2}\right)+5=
\begin{cases}
44.90,&\text{Bi$_2$Se$_3$},\\
44.61,&\text{Bi$_2$Te$_3$},
\end{cases}
\eeq
leading to a change of only $\frac{44.90}{44.61}\simeq1.0065,$ 
that is, approximately $0.7\%$.}} Corrections are perturbative if the radiative contributions in $m_{\textit{eff}}$ are less than $m_R$, fixing the range $g \leq 83.3\; {eV}^{-1}$. {Another connection with the previous discussion on the influence of the cutoff can be made by noticing that the higher limit of the coupling defining the perturbative regime diminishes, if it increases, according to the expression of the renormalized gap.}

 Therefore, a topological system induced by perturbative radiative corrections occurs in the parameter range
 \bea  3\times10^{-4} \;{eV}^{-1} \leq g\leq 83.3\;{eV}^{-1}. \eea
 \indent According to previous discussions, is worth mentioning that in the continuum limit, the condition for a topological phase compatible with the existence of the mentioned robust gapless boundary states, with spin and momentum locking, can be achieved,{if the momentum can take any arbitrarily large value,} through the topological Hamiltonian condition $ m_{\textit{eff}}(E=0)B>0 $, with $B={(3eg\ln{\frac{\Lambda^2}{m^2}}+5eg  )}/{32\pi^2}$ being the coefficient, up to the overall sign, of $\vec{p}^{\,2}$ term appearing in the energy-momentum dependent mass gap $M(p)$ \cite{shen,B1,B2}.\\ \indent{At this point, it is interesting to address the thin film idealized limit associated with the $L\to 0$ in our effective continuous field description. First, consider the expression for the BF propagator and the fermion one, in terms of a sum over the $n$ modes associated with the $p_3$ periodicity. At the aforementioned limit, the dominant expression for both propagators comes from the $n=0$ term. It implies an effective dimensional reduction, keeping all other remaining features. Therefore, the loop integrals in the lower dimension are finite, while the Dirac algebra in the numerator has the same overall structure.} {The latter reasoning implicitly considered the loop for periodic fields as a straightforward adaptation of the established formula expressed as a mode sum and an integral over continuous momenta \cite{kapusta}. For the zeroth mode, it reads}
\bea \!\!\!\!\!\!\!\!\!i\Sigma(p)\!=\!-\frac{ieg}{4L}\!\!\!\int\!\! \frac{d^3k}{(2\pi)^3}\frac{\gamma_\beta  \varepsilon^{\chi \beta \mu \nu}(p_\parallel \!-\!k)_\chi \gamma_\mu \gamma_\nu \gamma_5(m\!-\!\slashed{k})  }{(k-p_\parallel)^2[k^2-m^2]}.                    \eea
 {In this case, the sum of contributions for the momentum-dependent mass term arising from the self-energy at $p^2\approx m^2_R$, defining the analogy with the (TI) in the vicinity of the Dirac point, reads  }
 \bea \Sigma(p^2_\parallel \approx m^2_R)=\frac{3\ eg }{(4^{3/2}\pi\ L)}\sqrt{p^2_{\parallel}}+\cdots \eea
{It demonstrates that the resulting theory evaluated in that limit is not within the class of modified Dirac theories that possess non-trivial topological properties}. This feature is connected to the thin-width-limit analysis of the potential interaction presented in the next section. \\
\indent  Now, after the latter digression, let us return to the main discussion of this section, associated with the large width domain. Interestingly, according to Ref. \cite{shen}, even in the case of a gapped bulk topological insulator,  this kind of emerging renormalized structure with non-trivial topological properties derived here ensures gapless boundary modes with conical dispersions resembling massless Dirac particles. These modes also have a helical nature, displaying a correlation between their spin and momentum. {It is worth mentioning that throughout this article, the term boundary modes is a shorthand for topological solutions that live near the boundary planes. The solution is such that it vanishes at both boundaries but is concentrated near them \cite{lu2010, B2}.\\
\indent {Regarding the validity of our investigation when full finite-size effects are included, one can consider the following. The core piece of both bosonic and fermionic propagators can be alternatively written }as\footnote{For the case of a fermion field, the propagator core has a massive instead of a massless pole}
\begin{widetext}
\begin{eqnarray}
  \Delta (x_{\scalebox{0.6}{$\parallel$}},x_3)=\int \frac{d^4p}{(2\pi)^4}\frac{e^{-ip^\mu x_\mu}}{p^2}-\int \frac{i d^3p_{\parallel E} \ e^{-ip^{\mu}_{\parallel E} x_{\mu  \parallel}-\sqrt{p^2_{\parallel E}}|x_3|}}{(2\pi)^3\sqrt{p^2_{\parallel E}}\big(e^{2\sqrt{p^2_{\parallel E}}L}-1\big)},   \end{eqnarray}
  \end{widetext}
 {with one term recovering the propagator for the non-compactified space and the other encoding the finite size effects. The latter is suitably written in terms of an equivalent Wick-rotated integral formulation. Here, $p_{\parallel E}^\mu=(ip_{0E},\vec{p}_\parallel)$ and the definition $p^2_{\parallel E}\equiv -p_{\parallel E}^\mu p_{\mu  \parallel E}=p_{0E}^2+|\vec p_\parallel|^2$. Then, the fermion self-energy associated with propagators with such a structure is composed of two parts. An L-dependent one, which is finite, and a second piece recovering the previously obtained radiative correction for infinite volume. Both have different functional forms, with the latter being dependent on the physical cutoff. Therefore, for large enough samples, the topological number computation is still defined by the same previously calculated values unless $L$ is varied into a threshold scale at which the $L$-dependent contributions to such a number, if they include some non-trivial topological parts, achieve a critical magnitude. Since the magnitude of these finite-size contributions increases for narrower TI slabs, the possibility of topology-changing geometrical transitions should be associated with reducing the sample size relative to a larger-width configuration, in accordance with our previous discussions. Namely, for sufficiently thin films, the wave functions near both boundaries develop a considerable overlap, leading to a possible breaking of the topological character \cite{B2}.} {This interpretation is fully consistent with the effective continuous description adopted throughout this work, whose validity requires length scales much larger than the microscopic lattice spacing. As the slab thickness approaches the surface-state penetration length, the overlap between the opposite boundary wave functions becomes non-negligible, and the effective low-energy description must be supplemented by additional finite-size contributions. Consequently, the present framework naturally captures the field-theoretical mechanism underlying thickness-driven topological transitions. Material-dependent critical thicknesses are expected to depend on microscopic parameters that lie beyond the effective continuum description adopted here. }}

\section{Interparticle potential between boundary modes with topological protection \label{4}}

\begin{figure}[H]
	\centering
\includegraphics[width=8cm]{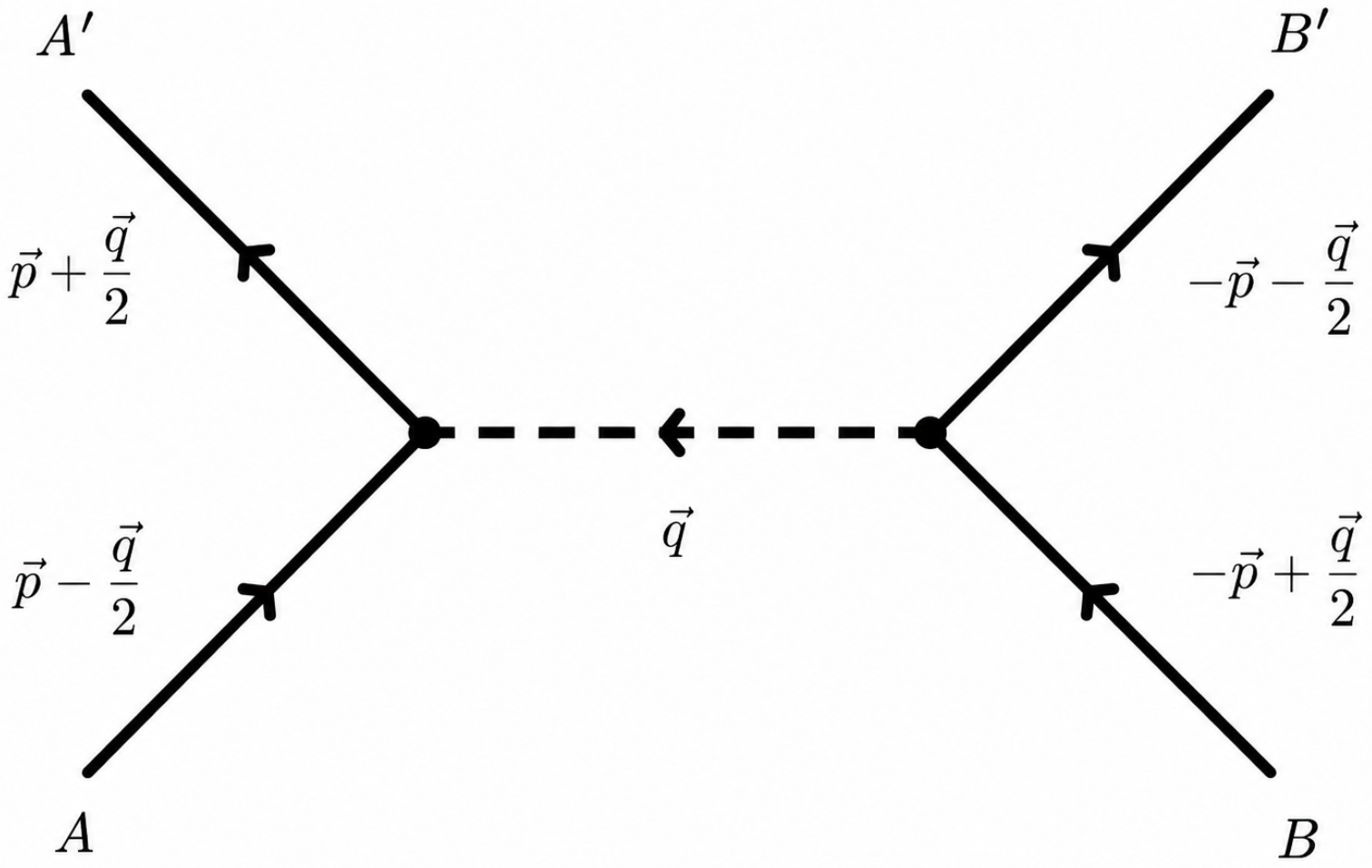}
	\caption{
Momentum assignment in the evaluation of one of the diagrams contributing to the tree-level scattering amplitude between boundary quasi-particles $A$ and $B$. In the center-of-mass frame, particle $A$ scatters from momentum $\vec p-\vec q/2$ to $\vec p+\vec q/2$, whereas particle $B$ scatters from $-\vec p+\vec q/2$ to $-\vec p-\vec q/2$. The dashed internal line represents the exchanged BF boson carrying momentum $\vec q$. It is worth highlighting that each vertex has an implicit different structure in this schematic representation.}
	\label{fig2}
\end{figure}

 \indent According to the last section, the renormalized fermionic structure satisfies the topological criterion associated with the existence of gapless helical boundary modes. The purpose of the present section is not to derive these modes from the interaction potential itself, but rather to investigate whether the effective tree-level interaction between boundary fermions mediated by the BF bosons exhibits correlations compatible with the expected topological boundary phenomenology. Then, we consider the case of states trapped near the boundaries with no overall momentum $p_3$ along  $x_3$. However, we admit a low-momentum exchange $q_3$ towards $x_3$ to include inter-boundary distance effects in the interaction potential. We keep the tree-level external states massive since the effective masslessness of the helical states is a feature of a specific solution of the renormalized fermion that already includes all one-loop interactions. 
{\indent Before deriving the effective interaction potential, let us explicitly state the regime of validity of the approximations employed throughout this section. We consider elastic scattering with weak coupling and low momentum transfer, such that the Born approximation provides the leading contribution to the scattering amplitude. Furthermore, we restrict our analysis to processes satisfying $|\vec q|\ll |\vec p|$, with external states corresponding to boundary quasi-particles. Therefore, the resulting potential should be interpreted as a leading-order effective interaction between boundary modes and not as a complete non-perturbative description of the topological phase.}  
 In this case, it can be obtained from the Fourier
transform of the amplitude associated with the quasi-particle scattering. We only consider the contribution from the direct scattering, since
the use of antisymmetric wave functions automatically
includes the exchange of identical fermions \cite{8, acci, gugu}.  Firstly, we analyze the system beyond the thin-film regime, with no compactification in the $x_3$ direction, achieved in the large $L$ limit. \\
\indent The amplitude for the interaction between two particles $A$ and $B$  in the center of mass frame yields 
 \begin{widetext}
\begin{eqnarray}
     \!\!\!\!\! i\mathcal{M}(\vec{q})=\Bigg\{  \Bar{\mathcal{U}}_{s_A}\!\left(\vec{p}_A\!+\!\frac{\vec{q}}{2}\right)\!(ie\gamma_\mu) \mathcal{U}_{s_A}\!\left(\vec{p}_A\!-\!\frac{\vec{q}}{2}\right)\frac{\varepsilon^{i\mu \nu \sigma}q_i}{\vec{q}^{\,2}}\Bar{\mathcal{U}}_{s_B}\!\left(\vec{p}_B\!-\!\frac{\vec{q}}{2}\right)\!\frac{ig}{4}[\gamma_\nu, \gamma_\sigma ]\mathcal{U}_{s_B}\!\left(\vec{p}_B\!+\!\frac{\vec{q}}{2}\right)\nonumber \\-
     \Bar{\mathcal{U}}_{s_B}\!\left( \vec p_B-\frac{\vec q}{2}\right)\!(ie\gamma_\mu) \mathcal{U}_{s_B}\!\left(\vec{p}_B\!+\!\frac{\vec{q}}{2}\right)\frac{\varepsilon^{i\mu \nu \sigma}q_i}{\vec{q}^{\,2}}\Bar{\mathcal{U}}_{s_A}\!\left(\vec{p}_A\!+\!\frac{\vec{q}}{2}\right)\!\frac{ig}{4}[\gamma_\nu, \gamma_\sigma ]\mathcal{U}_{s_A}\!\left(\vec{p}_A\!-\!\frac{\vec{q}}{2}\right)\Bigg\} \end{eqnarray}
     \end{widetext}
 {The amplitude is composed of a sum of two terms because both external fermions contribute to the current and also the spin source. Then, the momentum exchange in both terms has opposite sign.  Here, $\vec p_A=\vec p=-\vec{p}_B$ according to Figure \ref{fig2} and $s_A$ and $s_B$ denote the spin of the $A$ and $B$ particles, respectively.   }

The spinors  $\mathcal{U}_s$ describe free fermions with parameters constrained to emulate part of the typical TI  phenomenology, representing the external states of the amplitudes constructed from the theory's vertices \cite{gugu,helayel}. { The boundary helical states, associated with solutions of the renormalized fermion, display spin and momentum locking, and also absence of backscattering. Therefore, a representative tree-level potential for such a system should be associated with processes with no spin-flipping. Then, it defines an extra physically motivated constraint. }\\
\indent Here $|\vec{p}|$ represents the non-perturbed momentum modulus of each of these modes.  Owing to the low momentum transfer limit, we assume $\frac{|\vec{q}|^2}{|\vec{p}|^2}\approx 0$. The overall low energy setup furnishes $\frac{|\vec p|^2}{m^2}\approx 0$. Moreover,  energy conservation leads to $\vec p\cdot\vec q=0$.\\
\indent Therefore, considering the expressions for the sources in Appendix A in the low momentum transfer limit, one obtains an expression for the potential from the Fourier transform of the amplitude \cite{gugu,helayel}
\bea
    V(\vec{r})=-\int \frac{d^3q}{(2\pi)^3}e^{i\vec{q}\cdot\vec{r}}\mathcal{M}(\vec{q}), \eea
Integrating the $q_3$ momentum using the residue theorem leads to an expression compatible with eq.\eqref{prop2} in the ($L\to \infty$) limit, also including contributions from the sources. Therefore, considering the long-range structure, up to delta contact terms, the explicit form of the potential reads
\begin{widetext}
\begin{eqnarray}
V(\vec{r})=-eg\Bigg\{\frac{\langle \sigma_3^A\rangle \langle \sigma_3^B\rangle}{8\ m \pi}\left(-\frac{2}{r^3_{AB}}+\frac{3r^2_{AB \parallel}}{r^5_{AB}}\right)+\frac{2\varepsilon^{il3}r_{i\parallel}^{AB}p_l }{4\pi\ m  r^3_{AB}}\Big( \langle\sigma_3^A\rangle+ \langle\sigma_3^B\rangle\Big)   \Bigg\},      
\end{eqnarray}
\end{widetext}
  with the third component $r^{AB}_3=\pm 2L$, or equal to zero when referring to quasi-particles at the same surface. The form of the potential reveals the expected inter-boundary interaction suppression.  Here, the $x_3$ component of quasi-particle spin is given by $S_3=\frac{1 }{2}\langle\sigma_3^A\rangle$,  in natural units. 
  
  {The potential contains a term that does not depend on the spin sign and also a spin-orbit part. Interestingly, quasi-particles with different spins do not interact through the characteristic spin-orbit term.   Moreover, beyond the spin-orbit part, considering particles in the same boundary, states with the same spins also interact via an attractive central potential, whereas the ones with different spins repel each other.} \\
  \indent {{The spin-orbit contribution is proportional to
\begin{eqnarray}
(\langle \sigma_3^A\rangle+\langle \sigma_3^B\rangle)L_3,
\qquad
L_3=\epsilon^{il3}r_ip_l.
\end{eqnarray}
Therefore, the interaction energetically favors configurations in which the spin and orbital angular momentum are correlated. This structure is analogous to the standard spin-orbit coupling encountered in condensed matter systems. Although such a correlation does not uniquely determine the helical configuration, it is compatible with the spin-momentum locking expected for the topological boundary modes obtained from the renormalized fermionic sector.}  \\
\indent The force between quasi-particles with the same spin derived from this potential reads
\begin{widetext}
\begin{eqnarray}
     F^k&=&eg\Bigg\{\frac{\langle \sigma_3^A\rangle \langle \sigma_3^B\rangle}{8\ m \pi}\left[ \frac{6(r^k_{AB}+r^i_{AB}\delta_{\parallel \ i}^{k})}{r^5_{AB}}   -\frac{15\ r^k_{AB} r^2_{\scalebox{0.6}{$\parallel$}}}{r^7_{AB}}  \right]\nonumber \\ &&+\frac{2\varepsilon^{kl3}p_l }{4\pi \ m r^3_{AB} }\Big( \langle\sigma_3^A\rangle+ \langle\sigma_3^B\rangle\Big)-\frac{6r^k_{AB}\varepsilon^{il3}r_{iAB}p_l }{4\pi \ m r^5_{AB} }\Big( \langle\sigma_3^A\rangle+ \langle\sigma_3^B\rangle\Big) \Bigg\},                              \end{eqnarray}
\end{widetext}
    {Considering the case of an interaction between particles at the same boundary, there is a force orthogonal to the particle's momenta leading to an opposite axial torque exerted on each particle proportional to $\vec p\cdot\vec r_{AB}$, with the direction of the torque depending on the particle's spin.    Then, rotational equilibrium in the place of each particle demands the orthogonality relation $\vec p\cdot\vec r_{AB}=0$. However, this is too restrictive for the configurations. On the other hand, if the particles have the same momentum, which translates to $\vec p=0$ in the center-of-mass frame, it can be achieved. Therefore, the potential is such that no spin-orbit interaction occurs for particles with different spins. The ones that share the same quantum number approach the equilibrium when reaching equal momentum. Remarkably, this physically motivated tree-level interaction, even though it cannot establish the specific helical configuration, generates constraints that can include it.    }

\subsection{On the finite-width regime}
{Considering the propagator from eq. (\ref{prop2}), one can provide a formal expression for the complete potential based on the last calculations
\bea V_{\textsc{complete}}(\vec r)=\coth{\Big(L\sqrt{-\nabla^2_{\scalebox{0.6}{$\parallel$}}}\Big)}V(\vec r)\eea
The differential operator $-\nabla^2_{\scalebox{0.6}{$\parallel$}}$ has positive eigenvalues \cite{Hell:2023mph}. The expression reveals that the absence of spin-orbit correlation between modes with opposite spins is kept in the full expression. {Considering the Fourier representation, one can compute the finite-size effects using the Lipschitz-Hankel integral identity and an expression for $\coth(|\vec p_\parallel|L)$ written in terms of geometric series. Therefore, the spin-orbit (SO) part reads}
\begin{eqnarray}\label{ae1}\!\!\!\!\!\!\!\!\!\!\!\!V_{\textsc{complete}}^{{\rm (SO)}}(\vec r)=-eg\frac{\varepsilon^{il3}r_{i\parallel}^{AB}p_l }{2\pi\ m  }\Big( \langle\sigma_3^A\rangle+ \langle\sigma_3^B\rangle\Big)\nonumber \\ \!\!\!\!\!\!\times \Bigg(\frac{1}{{(r^2_\parallel\!+\!|r_3|^2)}^{3/2}}+2\sum_{n=1}^\infty \frac{1}{(r^2_\parallel\!+\!(2nL\!+\!|r_3|)^2)^{3/2}}  \Bigg)\end{eqnarray}
From Eq. (\ref{ae1}), the expected rotational symmetry breaking occurs, keeping just residual plane rotations. In the following, we derive the thin-film limit and point out some interesting physics in connection with the previous loop analysis.
\subsection{On the thin-film limit}

\indent Interestingly, in the thin-film limit, the correlations are stronger since the potential is enhanced by a factor $1/L$, and the interaction also decays more slowly with the inter-quasi-particle distance. It can break the topological order by an augmented inter-boundary coupling. Considering this limit, using the propagator from \eqref{prop3}, the potential explicitly reads\footnote{$V(\vec r)$ denotes the previously obtained inter-quasi-particle potential.} 
\begin{widetext}
\begin{eqnarray}
     V^{\textsc{thin}}(\vec{r}_{\scalebox{0.6}{$\parallel$}}, r_3)=-V(\vec{r}_{\scalebox{0.6}{$\parallel$}}, 0)\frac{|r_{3\,AB}|}{2L}+eg\left[ \frac{2\varepsilon^{ik3}r_i^{AB} }{2r^2_{ {\scalebox{0.6}{$\parallel$}} AB}}        \frac{p_k}{2\pi \ m L}(\langle\sigma_3^A\rangle+\langle\sigma_3^B\rangle)  \right]  \Bigg\}.   \end{eqnarray} 
     \end{widetext}
 Regarding interactions between quasi-particle pairs at different boundaries, for a planar distance projection $|\vec r_{\scalebox{0.6}{$\parallel$}}|=L $, the characteristic spin-orbit interaction potential associated with topological features is canceled. This feature indicates that the topological ansatz for boundary modes may not be an accurate description for the present limit. This analysis complements the conclusion of the last section describing the loss of topological order in the renormalized structure when considering the thin-film limit $L\to 0$. {It should be emphasized, however, that the present result is obtained within the effective long-distance approximation used throughout this section, including the expansion of the hyperbolic cotangent functions employed in the propagator analysis. Therefore, the cancellation of the spin-orbit contribution at $|\vec r_{\parallel}|=L$ should be interpreted conservatively. The present calculation establishes that the effective boundary-state description becomes unreliable in this regime, but it does not allow us to determine whether the cancellation survives in the exact theory. Consequently, we regard this result primarily as an indication of the breakdown of the effective topological phase in the thin-film limit rather than as a definitive prediction of a physical singular behaviour.} \\
\indent {It is worth mentioning analogous experimental observations. For HgTe quantum wells, samples thinner than the critical thickness $d_c$ behave as trivial insulators, whereas samples with $d>d_c$ realize the topological (quantum spin Hall) phase. Experimentally, the transition has been observed for thicknesses ranging approximately from $4.5$ nm to $12$ nm, with the critical value $d_c\simeq6.3$ nm \cite{shen,molen}. Similar thickness-driven topological transitions have also been reported for the three-dimensional topological insulators Bi$_2$Te$_3$ and Bi$_2$Se$_3$, where films with thicknesses ranging from one to ten quintuple layers (each approximately $1$ nm thick) exhibit a crossover associated with the increasing hybridization between the top and bottom surface states as the film thickness decreases \cite{quinze}.} {Our model can capture, using two different complementary approaches, the qualitative tendency toward the loss of topological order in the thin-film limit, consistently with the experimentally observed transitions. However, since the analysis is based on an effective long-distance description, it cannot determine the exact critical thickness associated with the phase transition. A field theory guess can be made by admitting that in real TI, the cutoff associated with the continuous description is related to a given scale $\Lambda \approx \frac{1}{L_{eff}}$ that can be much higher than the previously considered largest primitive vector size. In this case, in connection with the loop analysis, when the thin film limit ($L \to 0$) is considered, the $n=0$ approximation in loop propagators occurs when the energy cutoff becomes smaller than the energy scale $\frac{1}{L}$ occurring in the axial momentum expression, leading to the aforementioned dimensional reduction and loss of topological properties. Therefore, regarding the experimental examples cited in this subsection, the threshold energy scale would be associated with the critical thin film widths and the effective Fermi velocities.    }

\section{Conclusions \label{5}}
 \indent Throughout this article, we evaluated the features of the renormalized fermionic structure resulting from the interaction of ordinary fermions mediated by a BF bosonic sector. We explicitly demonstrated that it resembles the general structure of a three-dimensional  TI near the Dirac point. To implement the aforementioned coupling, a suitable set of auxiliary fields was included, ensuring that only the transverse sectors of the source tensors contribute to the physical processes. The analysis of the renormalized structure enabled the determination of a range of coupling parameters consistent with non-trivial topological properties, a necessary condition for the existence of gapless boundary modes.\\
 \indent Therefore, considering this setup, we evaluated the potential interaction between tree-level external boundary states in a compactified space defining the bulk region.  Finite-size effects were computed, and the transition to the thin-film limit was also investigated. Interestingly, a spin-orbit interaction, characteristic of this class of systems, arises. The equilibrium configuration of the associated force defines correlations that can include setups compatible with the so-called helical modes with spin and momentum locking. The complete solution, as well as the features associated with the phase transition to the thin film, was analysed and compared with analogous processes occurring in a set of real topological insulators.  \\
 
\appendix

\section*{Acknowledgements}

A. A. Nogueira thanks (PROEPD/UDESC) for full support. G. B. de Gracia thanks UFTM for hospitality and support. R. da Rocha thanks to The S\~ao Paulo Research Foundation (FAPESP) 
(Grants No. 2021/01089-1, No. 2025/23004-9, and No. 2024/05676-7) and the National Council for Scientific and Technological Development (CNPq) (Grants No. 303742/2023-2 and No. 401567/2023-0). 

\section{Source terms}

In order to incorporate the characteristic features of gapless charge-carrier excitations localized at the boundary, ensured by the renormalized fermionic response, we impose the physically motivated limit
\bea
p_3 \to 0,  \qquad 
\eea

The spinor solution reads \cite{gugu,helayel}
\bea
\mathcal{U}_s(p)
=\begin{pmatrix}
\eta_s \\
\dfrac{\,\vec{\sigma}\cdot\vec{p}}{E+m}\,\eta_s
\end{pmatrix},
\label{spinor_general}
\eea \\
where $\vec p\cdot\vec q=0$, from energy conservation,   $|\vec p|^2/m^2 \approx 0$ and $E\approx m$ due to low energy setup, and $|\vec q|^2/|\vec p|^2 \approx 0$ related to low momentum transfer. Here, $\eta_s$ are two-component spinors defined as
\bea
\eta_1 =
\begin{pmatrix}
1 \\ 0
\end{pmatrix},
\qquad
\eta_2 =
\begin{pmatrix}
0 \\ 1
\end{pmatrix}.
\eea
Implementing the previously defined limits, the general structure from eq.~(\ref{spinor_general}) provides an effective description of boundary excitations whose momentum component $p_3$  vanishes.

The spinors satisfy the identities
\begin{eqnarray}
\eta_s^\dagger \eta_r &=& \delta_{sr},\nonumber\\
\langle \sigma_i \rangle_s
&\equiv&
\eta_s^\dagger \sigma_i \eta_s
=
\langle \sigma_3 \rangle^s \delta_{i3}.
\end{eqnarray}
Under the above constraints, the fermionic sources acquire simplified forms. 
The vector current in our physically motivated setup of discarding spin-flipping processes is defined as
\bea
J_\mu^s
=
e\,\bar{\mathcal{U}}_s\!\left(p+\tfrac{q}{2}\right)
\gamma_\mu
\mathcal{U}_s\!\left(p-\tfrac{q}{2}\right),
\eea
yielding
\bea
J_0^s &=& e, \nonumber \\
J_i^s &=& \frac{e}{m}\left[
\!\ p_i 
-
\,\tfrac{i}{2}\,
\epsilon_{ijk} q_j \langle \sigma^k \rangle^s
\right].
\eea\\
The pseudo-tensor source is defined by
\bea
T_{\mu\nu}^s
=
\frac{g}{4}\,
\bar{\mathcal{U}}_s\!\left(p+\tfrac{q}{2}\right)
\gamma_5
\big[\gamma_\mu,\gamma_\nu\big]
\mathcal{U}_s\!\left(p-\tfrac{q}{2}\right),
\eea
from which we obtain
\bea
T_{0i}^s &=& \frac{g}{2}\,\langle \sigma_i \rangle^s, \nonumber \\
T_{ij}^s &=&
\frac{g}{2m}
\left(
p_i \langle \sigma_j \rangle^s
-
p_j \langle \sigma_i \rangle^s
\right)
+
\frac{g}{m}\,\frac{i}{4}\,\varepsilon_{ijk} q_k .
\eea


\begin{thebibliography}{40}
\bibitem{quatro} B. A. Volkov, and O. A. Pankratov, JETP Lett.  {\bf{42}}, 178 (1984).

\bibitem{um} L. Fu, C. L. Kane, and E. J. Mele, Phys. Rev. Lett. {\bf{98}}, 106803  (2007).
\bibitem{dois} L. Fu, and C. L. Kane, Phys. Rev. B {\bf{76}}, 045302 (2007).
\bibitem{tres} R. Roy, Phys. Rev. B {\bf{79}}, 195322 (2009).
\bibitem{cinco} D. Hsieh, D. Qian, L. Wray, Y. Xiu, Y. S. Hor, R. J. Cava, M. Z. Hasan, Nature {\bf{452}}, 970 (2008).
\bibitem{seis} M. Z. Hasan and J. E. Moore, Annu. Rev. Condens. Matter Phys. {\bf 2}, 55 (2011).
\bibitem{sete}H. Zhang, C. X. Liu, X.L. Qi, X. Dai, Z. Fang, and S. C. Zhang, Nat. Phys. {\bf{5}}, 438 (2009).
\bibitem{oito} Y. L. Chen, J. G. Analytis, J. H. Chu, Z. K. Liu, S. K. Mo, X. L. Qi, H. J. Zhng, D. H. Lu, X. Dai, Z. Fang, S. C. Zhang, I. R. Fischer, Z. Hussain, Z. K. Shen, Science, {\bf{325}}, 178 (2009).
\bibitem{nove} C.-X. Liu, X.-L. Qi, H. Zhang, X. Dai, Z. Fang, and S.-C. Zhang, Phys. Rev. B {\bf{82}}, 045122 (2010).
\bibitem{dez} K. Z. Zhang, H. V. Lepage, Y. Dong et al, Front. Phys. {\bf{19}}, 33208 (2024).
\bibitem{qb1} P. Fachin, F. Macheda, P. Barone, et al, Npj Comput. Mater. {\bf{11}}, 307,  (2025).
\bibitem{qb2} C. D. White and M. J. White, Phys. Rev. D {\bf{110}}, 11601 (2024).


\bibitem{onze} P. J. Rajput, S. U. Bhandavi, G.A. Wadha, Silicon {\bf{14}}, 9195 (2022).
\bibitem{doze} N. A. Niyazov, D. N. Aristov and V. Yu Kachorosvskii, Npj Comput. Mater.  {\bf{6}}, 174 (2020).
\bibitem{treze} S. P. Fluckey, S. Tiuari, C. L. Hinkle and W. G. Vanderberghe, Phys. Rev. Applied {\bf{18}}, 064037 (2022).
\bibitem{quatorze} M. Busch, O. Chiatti, S. Pezzini, et al, Sci. Rep. {\bf{8}}, 485 (2018).
\bibitem{1} D. Dudal, F. Matusalem, A. J. Mizher et al, Sci. Rep. {\bf{12}}, 5439 (2022).
\bibitem{3} D. Valenzuela, A. Raya and J. D. Garcia-Muñoz, Condens. Matter, {\bf{10}}, 1 (2025).
\bibitem{5} D. Liu, D. Sun, X. Hua, X. Jiang and N. Ma, Phys. Rev. D {\bf{108}}, 123022 (2023).
\bibitem{8} F. A. Barone, H. L. Oliveira and J. P. Ferreira, Phys. Rev. D {\bf{112}}, 065013  (2025).
\bibitem{9} M. Asorey and F. Esquerro, Phys. Rev. D {\bf{108}}, 045008 (2023).
\bibitem{10} F. Canfora, D. Dudal, T. Oosthuyse, P. Pais and L. Rosa, JHEP {\bf{2022}}, 95 (2022).

\bibitem{daRocha:2023waq}
R.~da Rocha, Annals Phys. \textbf{465}, 169663 (2024).


\bibitem{DaRocha:2020oju}
R.~da Rocha and A.~A.~Tomaz, J. Phys. A \textbf{53}, 465201 (2020). 

\bibitem{Ferreira-Martins:2019wym}
A.~J.~Ferreira-Martins, P.~Meert and R.~da Rocha, 
Eur. Phys. J. C \textbf{79},  646 (2019).

\bibitem{quinze}  Y. Zhang et al, Nat. Phys. {\bf{6}}, 584 (2010).



\bibitem{naka} N. Nakanish and I. Ojima, \textit{Covariant operator formalism of gauge theories and quantum gravity}, Singapore, World Scientific, 1990.
\bibitem{kr} G. B. de Gracia, Nucl. Phys. B {\bf{1001}}, 116498 (2024).
\bibitem{bf} A. Chan, T. L. Hughes, S. Ryu, E. Fradkin, Phys. Rev. B {\bf{87}}, 085132 (2013).

\bibitem{gab}G. B. de Gracia, B. M. Pimentel, and R. da Rocha, Annals Phys. {\bf{459}} 169545 (2023).
\bibitem{wang1} Z. Wang and B. Yan, J. Phys. Cond. Matter {\bf{25}}, 155601 (2013).

\bibitem{wang2}  Z. Wang and S. C. Zhang, Phys. Rev. X {\bf 2}, 031008 (2012).
\bibitem{shen} S. Q. Shen, \textit{Topological insulators; Dirac equation in condensed matter}, Springer-Verlag, Berlin, 2012.

\bibitem{B1}S. Q. Shen, W.-Y. Shan, H.-Z. Lu,           SPIN {\bf 1}, 33 (2011).

\bibitem{B2} \bibitem{shan2010}
W.-Y. Shan, H.-Z. Lu, and S.-Q. Shen, 
New J. Phys. \textbf{12}, 043048 (2010). 
\bibitem{Ozela:2021pse}
R.~F.~Ozela, V.~S.~Alves, G.~C.~Magalh{\~a}es and L.~O.~Nascimento, 
Phys. Rev. D \textbf{105},  056004 (2022).

\bibitem{kapusta} I. Kapusta, \emph{Finite Temperature Field Theory}, Cambridge University Press, Cambridge, 2017.

\bibitem{acci} A. Accioly, J. Heläyel-Neto, F. E. Barone, F. A. Barone, P. Gaete, Phys. Rev. D {\bf 90}, 105029 (2014).
\bibitem{gugu} G. P. de Brito, P. C. Malta and L. P. R. Ospedal, Phys. Rev. D {\bf{95}}, 016006 (2017).

\bibitem{helayel} F. A. Gomes Ferreira, P. C. Malta, L. P. R. Ospedal, J. A. Helayël-Neto  Eur. Phys. J. C {\bf{75}}, 232 (2015).

\bibitem{Hell:2023mph}
A.~Hell, D.~Lust and G.~Zoupanos, 
JHEP \textbf{02}, 039 (2024).


\bibitem{molen}  M. König, S. Wiedmann, C. Brüne, A. Roth, H. Buhmann, L. W. Molenkamp, X.L. Qi, S. C. Zhang, Science {\bf{318}}, 766  (2007).

\bibitem{lu2010}
H.-Z. Lu, W.-Y. Shan, W. Yao, Q. Niu, and S.-Q. Shen, Phys. Rev. B \textbf{81}, 115407 (2010).



\end{thebibliography}
\end{document}